\documentclass[prb,twocolumn,showpacs]{revtex4}
\usepackage{amsmath}
\usepackage{graphicx}
\usepackage{bm}

\newcommand{\be}{\begin{equation}}
\newcommand{\ee}{\end{equation}}

\newcommand{\tb}{Tb$_2$Ti$_2$O$_7$}

\newcommand{\Journal}[4]{{\em #1} \textbf{#2}, #3 (#4)}
\newcommand{\PRev}{Phys.\ Rev. }

\begin{document}

\title{Structural distortion and  the spin liquid state in \tb}

\author{S. H. Curnoe}
\email[Electronic address: ]{curnoe@physics.mun.ca}
\affiliation{Department of Physics and Physical Oceanography,
Memorial University of Newfoundland, St.\ John's, Newfoundland \& Labrador 
A1B 3X7, Canada}

\begin{abstract}
It is shown that a  $\vec{k}=0$, $A_{2u}$ distortion of the terbium tetrahedral network in
\tb\ causes the apparent isolation of single
tetrahedra as seen in neutron scattering studies.
Single tetrahedron collective spin states,
rather than individual
spins, account for the main features of the spin liquid state, namely,
fluctuating local moments and the absence of long range order.
Singlet and doublet collective spin ground states are considered.
An effective interaction between tetrahedra on the fcc lattice 
is derived and found to be weak and anisotropic.
\end{abstract}


\maketitle

%\section{Introduction}

\tb\ is an example of three dimensional
geometric frustration because
the magnetic Tb$^{3+}$ ions are arranged on a corner-sharing tetrahedral
network.
It has a spin liquid phase, characterised by the absence of long range
correlations and rapidly fluctuating local magnetic moments, which persists
down to at least 50 mK.\cite{gardner1999,gardner2003}
This article is devoted to another mysterious feature of \tb, namely
the apparent isolation of single tetrahedra from the tetrahedral network.
Such isolation is suggested by
neutron scattering experiments, which find that magnetic 
correlations beyond
the size of a single tetrahedron are absent,\cite{gardner1999}
and exact calculations of spin eigenstates on single tetrahedra
reproduce well diffuse neutron scattering patterns.\cite{curnoe2007,molavian2007}

A reasonable description of the
exchange interaction across the four Tb sites on a single tetrahedron 
in \tb\ has been
obtained only recently.\cite{curnoe2007,molavian2007}
The crystal electric field (CEF) ground state of the Tb ions is a doublet,
giving rise to a classical picture of Ising-like spins constrained to
point into or out of the four vertices of the tetrahedron.
However, mixing with higher CEF levels tends to
restore transverse spin components, negating the classical picture.
The resulting ground state resembles neither the ferromagnetic
``spin ice" states, which have two spins pointing inside the tetrahedron
and two pointing out,
%\cite{bramwell2001} 
nor the anti-ferromagnetic ``all-in or all-out" states.
Instead it is a linear combination of various states and with
no classical analogue.

The issue of how to deal with the entire tetrahedral network, and why the
single tetrahedron picture is valid, 
remains an outstanding problem.
Because of the connectivity of the tetrahedral network, eigenstates  of
a single tetrahedron are not in general 
eigenstates of the entire tetrahedral network.
This article offers a solution to this paradox,
which involves symmetry breaking in
the form of a lattice distortion.

Lattice distortions are often invoked in theory as a way to 
relieve geometrical frustration and arrive at an ordered 
state,\cite{yamashita2000}
and usually these involve a change of the crystal system,
for example, from cubic to tetragonal.
In fact, there is evidence that such
a transition occurs in \tb\ below any accessible temperature.\cite{ruff2007}
In  this article, we propose another kind of lattice distortion,
one which lowers the point group symmetry but leaves the crystal
system unchanged.
This distortion 
does not remove the frustration completely, but it
does serve to isolate single tetrahedra.
Collective spin states on single tetrahedra then replace individual
spins as
the fundamental basis for excitations and longer range
effective interactions.
The main features of the spin liquid
phase
follow from this scenario.

Pyrochlore crystals such as \tb\ belong to the cubic space group
$Fd\bar{3}m$ (O$_h$$^7$, No.\ 227), in which   
both Tb and Ti ions form separate,
corner-sharing tetrahedral networks. 
Here we are only concerned with the magnetic Tb$^{3+}$ ions.
The tetrahedra appear in two different orientations ($A$ and $B$), 
which alternate in 
the tetrahedral network. A tetrahedron of either type comprises
the primitive unit cell.  Thus the set of all $A$ tetrahedra 
form a
face-centred cubic (fcc) lattice, as do the set of all 
$B$ tetrahedra.  The set of $A$ tetrahedra are related to the set of
$B$ tetrahedra by the $\pi$ and $\pi/2$ screw rotation and inversion
elements of  $Fd\bar{3}m$.

We begin by considering the nearest neighbour isotropic (Heisenberg) exchange
interaction.  
The Hamiltonian for the entire tetrahedral network can be split into two terms
\begin{equation}
H_{ex} = {\cal J} H_A + {\cal J} H_B
\label{H}
\end{equation}
where ${\cal J}$ is the exchange coupling constant, 
\begin{equation}
H_A = \sum_k X_{(Ak)}
\end{equation}
and
$X_{(Ak)}$ is the exchange interaction over the $k$-th $A$ tetrahedron,
and $H_B$ is the exchange interaction over the  set of $B$ 
tetrahedra.  In either case, the sum over $k$ is a sum
over fcc lattice sites.  
The exchange interaction is
\begin{equation}
X = \sum_{<ij>} \vec{J}_i\cdot \vec{J}_j
\label{exchange}
\end{equation}
where the sum runs over four sites on a given tetrahedron.
The eigenstates of 
$H_A$ and $H_B$ are known, and
are simply the direct products across tetrahedra of the
eigenstates for a
single tetrahedron, which are given in Ref.\ \onlinecite{curnoe2007}
and described below.
However, in general, $[H_A,H_B] \neq 0$, therefore $H_A$ and $H_B$ do not
have common eigenstates with each other or with $H_{ex}$
(an exception to this is discussed below).
This is in apparent contradiction with the results of experiments,
which agree very well with the single tetrahedron picture.\cite{gardner1999,molavian2007,curnoe2007}
This suggests that  $H_{ex}$ (\ref{H})
should be replaced by an effective Hamiltonian\cite{mila}
\begin{equation}
H_{eff} = {\cal J}_A H_A + {\cal J}_B H_B
\label{effect}
\end{equation}
where one of the couplings is much larger than the other.
The difference between the exchange couplings ${\cal J}_A$ and ${\cal J}_B$ could
originate 
from a structural distortion which makes the $A$ tetrahedra smaller and
the $B$ tetrahedra larger, thus ${\cal J}_A > {\cal J}_B$, or vice versa.
Such a distortion is a $\vec{k}=0$, $A_{2u}$ mode of the Tb ions which 
reduces the space group symmetry from $Fd\bar{3}m$ to 
$F\bar{4}3m$.\cite{sergienko2003}   
This new space group is
compatible with an otherwise unexplained $(0,0,2)$ Bragg peak observed
in neutron scattering measurements.\cite{rule2006}  
If present, domains could produce isolated spins along their boundaries,
possibly contributing to observed glassy behaviour.\cite{luo2001,yasui2002,gardner2003}

Fits to the
dominant $(3,1,1)$, $(2,2,2)$, $(1,1,1)$ and $(2,2,0)$ Bragg peaks
are in good agreement with the pyrochlore space group $Fd\bar{3}m$.\cite{rule2006}  Therefore
it is reasonable to assume that the much smaller $(0,0,2)$ peak
is due to a small distortion of $Fd\bar{3}m$.  
The space groups which are compatible with all five of the peaks, are cubic, 
and which are also subgroups of $Fd\bar{3}m$ are 
$F\bar{4}3m$ (No.~216, T$_d^2$), $P\bar{4}3m$ (No.~215, T$^1_d$), $F23$ (No.~196, T$^2$), $P23$ (No.~195, T$^1$) and
$P2_13$ (No.~198, T$^4$).   
Among these, $F\bar{4}3m$ has the highest symmetry, and all are
subgroups of $F\bar{4}3m$.  
Thus it is strongly indicated that the space group of \tb\ is in fact 
$F\bar{4}3m$.
The $\vec{k}=0$, $A_{2u}$ displacement mode which gives rise to $F\bar{4}3m$ can occur
for the Tb, Ti or O ions,\cite{sergienko2003}
although physical considerations point in 
favour of Tb displacements, as discussed above, or O displacements due
to the role that the O ions play in mediating the exchange interaction.
%Thus it is strongly indicated that 
%that the symmetry operations which connect the set of $A$ tetrahedra to the set
%of $B$ tetrahedra are absent.
Whether or not the $(0,0,2)$ peak is present at all temperatures,
or appears due to some kind of exotic ``spin Jahn-Teller" effect, remains
to be investigated.\cite{bruce}

The ratio ${\cal J}_A/{\cal J}_B$ varies with the size of the lattice
distortion. 
The localised nature of the Tb $4f$ electrons ensures
that overlap integrals contributing to the exchange constants
will be very sensitive to changes in relative distances, and this
is evidenced by very large magnetostriction.\cite{alek1985}
Therefore it is possible that even a very small lattice distortion 
could produce a large difference between ${\cal J}_A$ and ${\cal J}_B$.
%which is required to validate the single tetrahedron picture.
%In fact, this seems to be the only way to 
%explain experimentally obtained diffuse neutron scattering
%patterns on \tb.\cite{gardner1999} 
%therefore we must be in the
%regime ${\cal J}_A \gg {\cal J}_B$ so that the second term in (\ref{effect})
%can be treated perturbatively with respect to the first.
%Finally, the model should also 
%include a secondary order parameter which appears in 
%the form of a small increase of lattice constant below
%18~K, coincident with the onset of spin liquid behaviour,\cite{ruff2007}
%and which will cause a decrease in both exchange couplings ${\cal J}_A$ and ${\cal J}_B$.

Assuming that $|{\cal J}_A| > |{\cal J}_B|$, 
we will find the ground state of $H_A$ and then consider 
$H_B$ as a perturbation.
In order to do this, we must first describe the collective spin states
of a single tetrahedron.  A tetrahedron
has four 
magnetic ions at its vertices, each of which has a local site symmetry of $D_{3d}$.  
Following the conventions established in Ref.\ \onlinecite{curnoe2007},
we label the ions \#1, 2, 3 and 4, where their $C_3$
axes point in the directions $[1,1,1]$, $[-1,-1,1]$, $[-1, 1, -1]$ and $[1,-1,-1]$
respectively.
Tb$^{3+}$ ions have a
total angular momentum of $J=6$, but the 13-fold degeneracy is split by   
the crystal electric field
into 5 singlets and 4 doublets; one of the doublets
is the ground state.  We take the ground state as\cite{gingras2000}
$|\pm\rangle = \pm0.13|\pm5\rangle\mp0.13|\mp1\rangle-0.95|\mp 4\rangle$, 
where the quantisation axis points in the direction of the $C_3$ axis for
each Tb ion.
There are 
then 16 collective spin states on a tetrahedron, which can be written
as $|\pm\pm\pm\pm\rangle = |\pm\rangle_1\otimes|\pm\rangle_2\otimes|\pm\rangle_3
\otimes|\pm\rangle_4$, where the subscripts indicate
the site on the tetrahedron.
%In general, the eigenstates of ${\cal J}X$ (Eq.\ \ref{exchange}) are 
%complicated linear combinations of these states.
Symmetry considerations predict that interactions
will split the  16 states into a singlet,
three doublets and three triplets,\cite{curnoe2007}
which are complicated linear combinations of the basis states
$|\pm\pm\pm\pm\rangle$.
%This {\em entanglement} is the source of paramagnetic behaviour as observed
%in $\mu$SR measurements.\cite{gardner1999}

The ground state spin configuration of Tb ions on a single tetrahedron for
isotropic anti-ferromagnetic exchange is
a doublet,\cite{curnoe2007} which we write as $|E_{\pm}\rangle$.
Otherwise the ground state depends on the details of the anisotropy.
An anisotropic interaction (equivalent to including
nearest neighbour dipole-dipole interactions)
was considered in Ref.\ \onlinecite{molavian2007}
and the singlet was found to be the ground state.   A triplet
ground state could also yield  patterns
similar to what are found in experiment.\cite{curnoe2007}
In the following, we will 
consider the doublet and the singlet as possible ground states on 
a single tetrahedron.

The ground state of $H_A$ is constructed by taking the direct product
of the single tetrahedron ground states across the $A$ tetrahedra.  
If the tetrahedron ground state is the singlet,
\begin{eqnarray}
&& |A_1\rangle = 
(|++--\rangle +|--++\rangle 
+|+-+-\rangle \nonumber  \\
&& +|-+-+\rangle 
 +|+--+\rangle +|-++-\rangle)/\sqrt{6}
\end{eqnarray}
then the ground state of $H_A$ can be written
$\Pi_k|A_{1}\rangle_k$, where $k$ indexes tetrahedra, that is, fcc lattice sites.  This state clearly has
long range correlations but they will be undetectable in 
static neutron 
scattering measurements, in agreement 
with experiment.\cite{gardner1999}  
This is because neutron scattering detects magnetic correlations,
which are proportional to
the matrix elements $\langle \Psi |\vec{J}_{i}|\Phi\rangle\langle\Phi|
\vec{J}_{j}|\Psi\rangle$, where $i$ and $j$ are Tb sites and
$\Psi$ and $\Phi$ are eigenstates of $H_A$,  which vanish unless
$i$ and $j$ belong to the same $A$ tetrahedron.
On the other hand, higher order correlations, beginning with quadrupolar,
will exist between tetrahedra.  Thus the state $\Pi_k|A_{1}\rangle_k$
is not a true spin liquid.
The excitation spectrum  of $H_A$ will 
be gapped, with an energy of the order of $0.1\sim1$~K,\cite{molavian2007} corresponding to the separation
between $|A_1\rangle$ and the first excited state on the tetrahedron.
%However, $|A_1\rangle$ does not 
%reproduce even qualitatively diffuse neutron scattering 
%patterns.\cite{curnoe2007}
%Apparently, in order to reproduce neutron scattering patterns it is necessary
%to include explicitly 
%higher crystal field levels in the wavefunction $|A_1\rangle$.\cite{molavian2007}

If the tetrahedron ground state is a doublet $|E_{\pm}\rangle$ then the 
ground states of  $H_A$ can be written as
$\prod_{k}|E_{\sigma}\rangle_k$. 
The ground state is highly degenerate.  One way of viewing the
degeneracy is to note that the state at  each site can be
any complex linear combination
of the doublet states $|E_{\pm}\rangle$.  
This SU(2) 
freedom on each fcc site leads to the absence of
correlations of any kind beyond the size of a single tetrahedron.
However, interactions coming from the 
perturbation $H_B$ will limit this freedom.
Ultimately, we find a system of weakly
interacting tetrahedra
arranged on an fcc lattice (which is also frustrated).
If the tetrahedron ground state is a triplet 
then there will be a SU(3) symmetry on each lattice site, which 
could also be 
limited by weak interactions.

In order to calculate the effect of the perturbation
$H_B$  we first need to examine in detail the exchange interaction
over a single tetrahedron and its eigenstates.  A useful representation
of the exchange interaction  over a single tetrahedron is
given in Ref.\ \onlinecite{curnoe2007} in which the angular
momentum operators are expressed in terms of local coordinate axes 
(indicated by subscripts) such that
the local $z$ axis for each Tb ion points in the direction of its $C_3$
axis.   Local $x$ and $y$ axes have also been implicitly selected.  
The result is
written in Table I.  
Only local coordinate operators are used in the following.
%Local coordinate operators are indicated by %$x$, $y$ or $z$ 
%subscripts, global
%operators by  superscripts.
The symmetry of the crystal
is octahedral, which permits three separate invariant
terms in the exchange interaction,
${\cal J}_1 X_1$, ${\cal J}_2X_2$ and ${\cal J}_3X_3$,
where ${\cal J}_{1,2,3}$ are different exchange couplings 
for each term, and $X_{1,2,3}$ are the sums of all the terms 
in the second, third and fourth columns respectively of Table I.
%CHANGE
There is actually a fourth invariant allowed under $O_h$ symmetry 
($X_3$ is split), which we do not consider here,
although the following discussion can be easily generalised.
The isotropic exchange interaction is the sum of all three terms
with ${\cal J}_1={\cal J}_2 ={\cal J}_3$.
In the following,
we will assume that the exchange interaction is 
anti-ferromagnetic.
%however we note that the matrix elements for the $J_{\pm}$ operators
%appear to be
%larger that those of the $J_z$ operators,\cite{curnoe2007,molavian2007}
%so that the exchange interaction can be approximated by ${\cal J}X_3$.

\begin{table*}[htb]
\begin{tabular}{|l||c|c|c||c|}
\hline
term  & $X_1$ & $X_2$ & $X_3$ & bond  \\ \hline
$\vec{J}_1\cdot\vec{J}_2$
& $-\frac{1}{3}J_{1z}J_{2z}$ & $- \frac{\sqrt{2}}{3}
[J_{1z}(J_{2+}+J_{2-})+(J_{1+}+J_{1-})J_{2z}]$&
$\frac{1}{3}(J_{1+}J_{2+}+J_{1-}J_{2-}) 
-\frac{1}{6}(J_{1+}J_{2-}+J_{1-}J_{2+})$ & $(1/2,1/2,0)$ \\
$\vec{J}_3\cdot\vec{J}_4$
& $-\frac{1}{3}J_{3z}J_{4z}$ & $- \frac{\sqrt{2}}{3}
[J_{3z}(J_{4+}+J_{4-})+(J_{3+}+J_{3-})J_{4z}]$&
$\frac{1}{3}(J_{3+}J_{4+}+J_{3-}J_{4-}) 
-\frac{1}{6}(J_{3+}J_{4-}+J_{3-}J_{4+})$ &$(-1/2,1,2,0)$ \\
$\vec{J}_1\cdot\vec{J}_3$ &
$-\frac{1}{3}J_{1z}J_{3z}$ &
$ - \frac{\sqrt{2}}{3}
[J_{1z}(\varepsilon J_{3+}+\varepsilon^2 J_{3-})+(\varepsilon J_{1+}+\varepsilon^2 J_{1-})J_{3z}]$ &
$ \frac{1}{3}(\varepsilon^2 J_{1+}J_{3+}+\varepsilon J_{1-}J_{3-}) 
 -\frac{1}{6}(J_{1+}J_{3-}+J_{1-}J_{3+})$& $(1/2,0,1/2)$ \\
$\vec{J}_2\cdot\vec{J}_4$ &
$-\frac{1}{3}J_{2z}J_{4z}$ &
$ - \frac{\sqrt{2}}{3}
[J_{2z}(\varepsilon J_{4+}+\varepsilon^2 J_{4-})+(\varepsilon J_{2+}+\varepsilon^2 J_{2-})J_{4z}]$ &
$ \frac{1}{3}(\varepsilon^2 J_{2+}J_{4+}+\varepsilon J_{2-}J_{4-}) 
 -\frac{1}{6}(J_{2+}J_{4-}+J_{2-}J_{4+})$ &$(-1/2,0,1/2)$ \\
$\vec{J}_1\cdot\vec{J}_4$ &
$-\frac{1}{3}J_{1z}J_{4z}$ &
$- \frac{\sqrt{2}}{3}
[J_{1z}(\varepsilon^2 J_{4+}+\varepsilon J_{4-})+(\varepsilon^2 J_{1+}+\varepsilon J_{1-})J_{4z}]$ &
$ \frac{1}{3}(\varepsilon J_{1+}J_{4+}+\varepsilon^2 J_{1-}J_{4-}) 
 -\frac{1}{6}(J_{1+}J_{4-}+J_{1-}J_{4+})$ & $(0,1/2,1/2)$ \\
$\vec{J}_2\cdot\vec{J}_3$ &
$-\frac{1}{3}J_{2z}J_{3z}$ &
$- \frac{\sqrt{2}}{3}
[J_{2z}(\varepsilon^2 J_{3+}+\varepsilon J_{3-})+(\varepsilon^2 J_{2+}+\varepsilon J_{2-})J_{3z}]$ &
$ \frac{1}{3}(\varepsilon J_{2+}J_{3+}+\varepsilon^2 J_{2-}J_{3-}) 
 -\frac{1}{6}(J_{2+}J_{3-}+J_{2-}J_{3+})$ & $(0,-1/2,1/2)$ \\
\hline
\end{tabular}
\caption{\label{patterns} The exchange interaction over
a single tetrahedron expressed in terms of local
coordinates for each Tb ion. The first column lists the terms 
in the exchange interaction over a single tetrahedron.  In each row,
the first entry is the sum of the middle three.  
The isotropic exchange is the sum of all the terms in the middle three
columns, and may be divided into three anisotropic terms $X_1$, $X_2$ and
$X_3$ which are sum of all the terms in each of the middle three columns. 
The last column lists the bond direction for each term.
$\varepsilon = \exp \frac{2 i \pi}{3}$.
}
\end{table*}

The matrix elements for $J_{z,\pm}$ are
\begin{eqnarray}
\langle \pm | J_z |\pm\rangle & = & \pm j\\
\langle \pm |J_{\pm} |\mp \rangle & = & t
\end{eqnarray}
The parameters $j$ and $t$ serve as a very useful 
characterisation of single-ion doublet spin states.
%As discussed later,  $t= 0$ represents a special case when $H_{ex}$ can be exactly solved.
For $J=1/2$ ions, $j=1/2$ and $t=1$, while for other $1/2$-integral
spins $j$ and $t$ may be quite different.
The parameter $t$ equals zero for integral spins (as for the Tb$^{3+}$ ion) but
may acquire a  significant non-zero value due to mixing 
with higher crystal electric field levels.\cite{molavian2007}
%In contrast, the crystal electric field splitting in the spin ice material Ho$_2$Ti$_2$O$_7$ ($J=8$) 
%is much greater, so $t\approx 0$ in that case. 
%In another spin ice
%Dy$_2$Ti$_2$O$_7$ ($J=??$) the caluculated values of $j$ and $t$ are
%respectively, and again a large crystal electric field ensures
%that $t$ does not get renormalised to a higher value.
A comparison between theoretical and experimental diffuse 
neutron scattering patterns
suggests that $t$ is larger than $j$ in \tb.\cite{curnoe2007,molavian2007}
Thus, in our initial approximation, we will assume that $j=0$; then only
term in the exchange interaction with non-zero matrix elements is
$X_3$.
%This leads us to perturbative expansions in two, probably related, variables:
%$J_B/J_A$ and $j/t$.

The ground state of $X_3$ is the doublet\cite{curnoe2007,note1}
\begin{equation}
|E_{\pm}\rangle = \sqrt{3/5}|E_{\pm}^{(1)}\rangle-\sqrt{2/5}|E_{\pm}^{(3)}\rangle
\end{equation}
where
\begin{eqnarray}
&&|E_{+}^{(1)}\rangle  =  |++++\rangle, \hspace{.2in}
|E_{-}^{(1)}\rangle  =  |----\rangle \\
&&|E_{+}^{(3)}\rangle  =  (|++--\rangle
+\varepsilon|+-+-\rangle +\varepsilon^2|+--+\rangle \nonumber \\
  && + |--++\rangle + \varepsilon|-+-+\rangle + \varepsilon^2|-++-\rangle)/\sqrt{6} \\
& &|E_{-}^{(3)}\rangle  = [|E_{+}^{(3)}\rangle]^{*}
\end{eqnarray}
where
$\varepsilon = \exp \frac{2 i \pi}{3}$.
%As described above, the direct product of $|E_{\pm}\rangle$ across all of the
%$A$ tetrahedra gives the ground states of $H_A$. 
%which we will call $|\Psi_{Ap}\rangle$, where $p$ indexes the degenerate ground states.
First order corrections due to $H_B$ are found by
calculating the matrix elements of $H_B$ between the degenerate
ground states of $H_A$.
We will express the final result as an effective interaction
between the ground state doublets on neighbouring tetrahedra.
%$\langle \Psi^{0}_{Ap'}|H_B|\Psi_{Ap}\rangle$.
Recall that $H_B$ is the exchange interaction summed over all $B$ tetrahedra.
It consists of terms of the form $\vec{J}_i\cdot\vec{J}_j$, where $i$ and $j$
are nearest neighbour but are found on different, neighbouring
$A$ tetrahedra.
Nearest neighbour $A$ tetrahedra share exactly one pair of nearest neighbour
magnetic ions.
Considering only the $X_3$ part of the exchange interaction, we find that
{\em all} of the ground state matrix elements 
%$\langle \Psi^0_{Ap'}|H_B|\Psi^0_{Ap}\rangle$ 
vanish\cite{note2}
because each term in $H_B$ raises or lowers at most one site in any $A$
tetrahedron, and the result is orthogonal to all of the
ground states.
Therefore, to  first order in ${\cal J}_B/{\cal J}_A$ and
zeroth order in $j/t$, the  $A$ tetrahedra are
non-interacting.

Now we consider finite values of  $j/t$
by including $X_1$ and $X_2$ in the exchange interaction.  
%We will assume
%that the exchange interaction is isotropic and proportional
%to $X_1+X_2+X_3$.
This
will add mixtures of 
\begin{eqnarray}
|E^{(2)}_{+}\rangle & = & (|+---\rangle +|-+--\rangle \nonumber\\
& & 
+|--+-\rangle + |---+\rangle)/2 \\
|E^{(2)}_{-}\rangle & = & (|-+++\rangle + |+-+++\rangle \nonumber \\
& & 
+|++-+\rangle + |+++-\rangle )/2
\end{eqnarray}
and
\begin{equation}
|E^{(3')}_{\pm}\rangle  = \sqrt{2/5}|E_{\pm}^{(1)}\rangle + \sqrt{3/5}|E_{\pm}^{(3)}\rangle
\end{equation}
to $|E_{\pm}\rangle$.
The general form is $|E_{\pm}\rangle = \alpha|E_{\pm}^{(1)}\rangle
+\beta|E_{\pm}^{(2)}\rangle + \delta|E_{\pm}^{(3)}\rangle$. 
%where $-\alpha/\beta \ge 1/\sqrt 2$.  
%For the special case of $J=1/2$, $\alpha = -\sqrt{2}/3$, $\beta = 2/3$
%and $\delta = 1/\sqrt{3}$.
To find the effective interaction between tetrahedra,
consider two neighbouring $A$ tetrahedra, with Tb sites numbering
1, 2, 3, and 4 on the first tetrahedron and 5, 6, 7 and 8 on the second. 
Projecting all possible ground states of $H_A$ onto the subspace
of these two tetrahedra yields four states of the form
$|E_{\pm};E_{\pm}\rangle$.  
Suppose that sites 
1 and 6 are nearest neighbours and that the $C_3$ axis of site 6 points in the same direction as that of site 2.  Then
$\vec{J}_1\cdot \vec{J}_6$ is a term in  $H_B$
and takes the same form as $\vec{J}_1\cdot \vec{J}_2$.
The effective interaction between tetrahedra is found by calculating the
16 matrix elements $\langle E_{\pm};E_{\pm}|\vec{J}_1\cdot \vec{J}_6|E_{\pm};E_{\pm}\rangle$.
The non-zero matrix elements are
%\begin{eqnarray}
%\langle E_{+};E_{+}|X_1| E_{+};E_{+}\rangle & = & 
%\langle E_{-};E_{-}|X_1| E_{-};E_{-}\rangle = -\gamma^2/3 \nonumber \\
%\langle E_{+};E_{-}|X_1| E_{+};E_{-}\rangle & = & 
%\langle E_{-};E_{+}|X_1| E_{-};E_{+}\rangle = \gamma^2/3 \nonumber \\
%\langle E_{+};E_{\pm}|X_2| E_{+};E_{\mp}\rangle & = & 
%\langle E_{\pm};E_{+}|X_2| E_{\mp};E_{+}\rangle = -\gamma \tau \sqrt{2}/3 \nonumber \\
%\langle E_{-};E_{\pm}|X_2| E_{-};E_{\mp}\rangle & = & 
%\langle E_{\pm};E_{-}|X_2| E_{\mp};E_{-}\rangle = \gamma \tau \sqrt{2}/3 \nonumber \\
%\langle E_{+};E_{+}|X_3| E_{-};E_{-}\rangle & = & 
%\langle E_{-};E_{-}|X_3| E_{+};E_{+}\rangle = \tau^2/3 \nonumber \\
%\langle E_{+};E_{-}|X_3| E_{-};E_{+}\rangle & = & 
%\langle E_{-};E_{+}|X_3| E_{+};E_{-}\rangle = -\tau^2/6 \nonumber
%\end{eqnarray}
\begin{eqnarray}
\langle E_{+};E_{\pm}|X_1| E_{+};E_{\pm}\rangle & = & 
\langle E_{-};E_{\mp}|X_1| E_{-};E_{\mp}\rangle = \mp\gamma^2/3 \nonumber \\
\langle E_{+};E_{\pm}|X_2| E_{+};E_{\mp}\rangle & = & 
\langle E_{\pm};E_{+}|X_2| E_{\mp};E_{+}\rangle = -\gamma \tau \sqrt{2}/3 \nonumber \\ 
\langle E_{-};E_{\pm}|X_2| E_{-};E_{\mp}\rangle & = & 
\langle E_{\pm};E_{-}|X_2| E_{\mp};E_{-}\rangle = \gamma \tau \sqrt{2}/3 \nonumber \\
\langle E_{+};E_{+}|X_3| E_{-};E_{-}\rangle & = & 
\langle E_{-};E_{-}|X_3| E_{+};E_{+}\rangle = \tau^2/3 \nonumber \\
\langle E_{+};E_{-}|X_3| E_{-};E_{+}\rangle & = & 
\langle E_{-};E_{+}|X_3| E_{+};E_{-}\rangle = -\tau^2/6 \nonumber
\end{eqnarray}
where
$\gamma = j(\alpha^2 - \beta^2/2)$ and $\tau = t \alpha\beta$.
From these matrix elements we can infer that the effective interaction
between tetrahedra is anisotropic in general and can be divided into
three separate terms 
\begin{equation}
H_{tetra} = {\cal J'}_1 X_1 +{\cal J'}_2 X_2 + {\cal J'}_3X_3,
\label{Htetra}
\end{equation}
where  $X_i$ now operate on tetrahedral ground state doublets, as follows.
There are four invariant operators on the fcc lattice that
%CHANGE
are directly related to the four invariants on the single tetrahedron
(note that, as explained above, only three have been considered here).
The bond direction between 
a tetrahedron and each of its 
twelve 
%CHANGE
nearest neighbours 
can be defined by the bond direction between the
nearest neighbour ion sites on
neighbouring tetrahedra.  Six different bond directions are
present.  The bond direction determines the
correspondence between rows in Table I and
terms in the effective interaction between tetrahedra. 
Assuming that
matrix elements of the tetrahedron operators 
are $\langle E_{\pm} |J_z|E_{\pm}\rangle = \pm 1$
and  $\langle E_{\pm}|J_{\pm}|E_{\mp}\rangle = 1$, we find
${\cal J'}_1 = {\cal J}_1 \gamma^2$, ${\cal J'}_2 = {\cal J}_2 \gamma\tau$
and ${\cal J'}_3 = {\cal J}_3 \tau^2$.
Higher order corrections  arising from mixing via $H_B$ between the 
single tetrahedron ground states
$|E_{\pm}\rangle$ and excited states will further
renormalise  ${\cal J'}_1$,
${\cal J'}_2$ and ${\cal J'}_3$.
The renormalisation of the coupling constants in Eq.\ \ref{Htetra} implies that
even if the underlying exchange interaction is isotropic, the effective
exchange interaction between tetrahedra is anisotropic.
In a similar fashion, the effective interaction between tetrahedra
with  triply degenerate ground states can also be found.

Anisotropy in $H_{tetra}$ may assist long range 
ordering of the tetrahedra. 
However, at temperatures so far obtained, there is no 
evidence from neutron scattering for any kind of inter-tetrahedra correlations.
The inelastic neutron scattering function is proportional to\cite{jensen}
\begin{widetext}
\begin{equation}
I({\bf q}) \propto \sum_{m}e^{-E_m/k_BT}
\sum_{i,j}\sum_{a,b}
\sum_{n}
(\delta^{ij}-\hat{q}^i\hat{q}^j)
\langle m|J_a^i|n\rangle\langle n|J_b^j|m\rangle e^{i{\bf q}\cdot ({\bf r}_b
-{\bf r}_a)},
\label{fm(q)}
\end{equation}
where $a$ and $b$ are the four magnetic ion 
sites at the corners of a tetrahedron and
$n$ and $m$ are eigenstates of the system.
The angular momentum operators $J^{i}$ refer to the global
coordinate system (indicated by superscripts).
This
formula yields patterns which 
are linear combinations of the following functions 
\begin{eqnarray}
f^1(h,k,l) & = & \frac{1}{h^2+k^2+l^2}
\left[hk \sin \frac{\pi h}{2} \sin \frac{\pi k}{2} +
kl \sin \frac{\pi k}{2} \sin \frac{\pi l}{2} +
hl \sin \frac{\pi h}{2} \sin \frac{\pi l}{2}\right] \\
f^2(h,k,l) & = &  \frac{1}{h^2+k^2+l^2}
\left[ h^2 \cos\frac{\pi k}{2}\cos\frac{\pi l}{2}
+ k^2 \cos\frac{\pi h}{2}\cos\frac{\pi l}{2} 
+ l^2 \cos\frac{\pi h}{2}\cos\frac{\pi k}{2}\right]\\
f^3(h,k,l) & = & \frac{1}{h^2+k^2+l^2}
\left[ h^2 \cos \frac{\pi h}{2}\left(\cos \frac{\pi k}{2}
+\cos\frac{\pi l}{2}\right) + k^2\cos\frac{\pi k}{2}
\left(\cos\frac{\pi h}{2} + \cos \frac{\pi l}{2}\right)
+l^2\cos\frac{\pi l}{2}\left(\cos\frac{\pi h}{2}+\cos\frac{\pi k}{2}\right)
\right].
\end{eqnarray}
\end{widetext}
Inter-tetrahedron correlations would be manifested in the appearance
of smaller period (in $k$ space)  contributions to the scattering patterns.  If
present, these could help to constrain the effective interaction
between tetrahedra.
However, their absence indicates that inter-tetrahedra interactions are 
indeed weak.

%Alternatively, a glassy or ice-like  (in the sense that the tetrahedra ``moments" get
%pinned to particular orientations) state may also develop.
%As in the $|A_1\rangle$ ground state, an energy gap of the order of $0.1\sim 1$ K
%is expected due excitations between the doublet $|E_{\pm}\rangle$ and
%the next excited state on a single tetrahedron.  Such a low energy excitation
%(1.4 K)
%has been reported very recently.\cite{mirebeau2007}

Finally, let us contrast our results to 
situations in which the commutator
$[H_A,H_B]$ does vanish, allowing common eigenstates of 
$H_A$, $H_B$ and $H_{ex}$.  
Non-zero terms in the commutator $[H_A,H_B]$ arise 
from terms proportional to 
$X_2$ and $X_3$ in $H_A$ and $H_B$.
Thus, if the exchange interaction
is highly anisotropic and only $X_1$ appears
{\em or} when the parameter $t$ 
vanishes, then $[H_A,H_B]=0$, and the eigenstates of
$H$, $H_A$ or $H_B$ are the sixteen basis states $|\pm\pm\pm\pm\rangle$.
If the sign of ${\cal J}_1$ is positive then the ground states of a
single tetrahedron are  the antiferromagnetic states
$|++++\rangle$ and $|----\rangle$; otherwise the ground state has a 
six fold degeneracy, $|++--\rangle$, $|--++\rangle$, $|+-+-\rangle$,
$|-+-+\rangle$, $|+--+\rangle$ and $|-++-\rangle$.
These situations can each be defined by rules:  ``all in or all  out"
in the former and ``two in/two out" (spin ice rule) in the latter.
Ground states of $H_A$ are again found by taking the direct product of
ground states over the $A$ tetrahedra.  A ground state of $H_A$ will be
an eigenstate of $H_B$ but in general it will not be a ground state 
unless the appropriate ground state rule is satisfied on all of the 
$B$ tetrahedra too.
The distinguishing feature between the special case when $[H_A,H_B]=0$ and
the general case is the form that the eigenstates take.  When the
commutator is non-zero the eigenstates of the single tetrahedron
 are necessarily entangled, that is, they must be linear
combinations of the basis states $|\pm\pm\pm\pm\rangle$, which
leads to fluctuating local moments found in \tb.
%reminiscent of Anderson's resonating valence bond.\cite{anderson1973}

To summarise, we have shown that a  $\vec{k} =0$, $A_{2u}$ lattice distortion
can account for the observed single tetrahedron behaviour.  
This distortion results in a space group of $F\bar{4}3m$, 
which is compatible with 
with a $(0,0,2)$ Bragg peak
observed by neutron scattering\cite{rule2006}
and the loss of inversion centre recently observed in Raman
scattering.\cite{lummen}
The eigenstates of the system are then direct products over the fcc
lattice of 
the single tetrahedron eigenstates.  
The effective interaction between 
tetrahedra 
is weak and anisotropic.

%\section{Discussion and Summary}

\begin{acknowledgments}
I thank E. Altman and M. Gingras for very helpful discussions. 
I am very grateful
to the Department of Condensed Matter Physics at the Weizmann Institute of
Science, especially A. Finkel'stein, for hospitality and support 
while this work was being completed, and also to the Shrum Foundation and NSERC
for support.
\end{acknowledgments}


\begin{thebibliography}{00}
\bibitem{note1} The singlet state $|A_1\rangle$ is also
a ground state of ${\cal J} X$ when $j=0$, but has a higher energy when $j\neq 0$.
\bibitem{bruce} The $(2,0,0)$ peak is likely present for
several other rare-earth titinates, including the spin ice Ho$_2$Ti$_2$O$_7$,
which implies that it is not due to coupling with the spin. (B. Gaulin, private
communication).
\end{thebibliography}
\end{document}